\documentclass[twocolumn,prb,10pt]{revtex4-1}
\usepackage{amsmath}
\usepackage{graphicx}
\usepackage{subfigure}
\usepackage{hyperref}
\usepackage{bm}

\begin{document}

\title{Topological thermoelectric effects in spin-orbit coupled electron and
hole doped semiconductors}
\author{E. Dumitrescu$^{1}$}
\author{Chuanwei Zhang$^{2}$}
\author{D. C. Marinescu$^{1}$}
\author{Sumanta Tewari$^{1}$}
\pacs{74.78.-w, 03.67.Lx, 71.10.Pm, 74.45.+c}

\begin{abstract}
We compute the intrinsic contributions to the Berry-phase mediated anomalous
Hall and Nernst effects in electron- and hole-doped semiconductors in the
presence of an in-plane magnetic field as well as Rashba and Dresselhaus
spin orbit couplings. For both systems we find that the regime of chemical
potential which supports the topological superconducting state in the
presence of superconducting proximity effect can be characterized by
plateaus in the topological Hall and Nernst coefficients flanked by
well-defined peaks marking the emergence of the topological regime. The plateaus arise from
a clear momentum space separation
between the region where the Berry curvature is peaked (at the
`near-band-degeneracy' points) and the
region where the single (or odd number of) Fermi surface lies in the Brillouin zone. The
plateau for the Nernst coefficient is at vanishing magnitudes surrounded by two
large peaks of opposite signs as a function of the chemical potential.
These results could be useful for experimentally deducing the chemical
potential regime suitable for realizing topological  states in the presence of
proximity effect.
\end{abstract}
\affiliation{
$^{1}$Department of Physics and Astronomy, Clemson University, Clemson, SC
29634 USA\\
$^{2}$Department of Physics and Astronomy, Washington State University,
Pullman, WA 99164 USA}
\maketitle

\section{Introduction}

Topological superconducting states with or without broken time-reversal
symmetry \cite{Schnyder,Kitaev-AIP,Ruy} have recently come under increasing
attention because of the possibility of realizing Majorana fermions.
Majorana fermions,\cite%
{Majorana,Wilczek,Nayak-Wilczek,Read-Green,Ivanov,Stern,Kitaev-QC,Nayak-RMP}
defined by self-hermitian second quantized operators $\gamma =\gamma
^{\dagger }$, are remarkable quantum mechanical particles which can be
construed as their own anti-particles. Recently it has been shown that a 2D
electron-doped semiconductor with Rashba type spin-orbit coupling in
proximity to a bulk $s$-wave superconductor and an externally induced
perpendicular Zeeman splitting can support a topological superconducting
phase with Majorana fermion modes at vortex cores and sample edges. \cite%
{Sau1,Index-Theorem,Long-PRB} The proposal followed on an earlier similar
proposal for topological superconducting states using spin-orbit coupling
and Zeeman fields made in the context of cold fermions. \cite%
{Zhang-Tewari,Sato-Fujimoto} It has also been pointed out by Alicea \cite%
{Alicea-Tunable} that in the presence of Dresselhaus spin-orbit coupling
co-existing along with the usual Rashba coupling, the topological
superconducting state in electron-doped semiconductors can be realized with
an in-plane Zeeman field. Since the in-plane Zeeman field can be directly
applied by an in-plane magnetic field which is free of the orbital effects,
the geometry proposed by Alicea is useful for producing topological
superconductivity in two-dimensional electron systems. Very recently it has
also been shown \cite{Mao} that the generic Luttinger Hamiltonian applicable
to the hole-doped semiconductors also supports topological superconductivity
and Majorana fermions in the presence of a perpendicular Zeeman field in a
manner similar to its electron-doped counterpart. The possibility of a
larger value of the effective mass and spin-orbit coupling in $p$-type holes
in a semiconductor quantum well makes the hole-doped systems an attractive
candidate for realizing topological superconductivity that breaks
time-reversal symmetry. The strategy of producing topological
superconductivity and Majorana fermions using Rashba spin-orbit coupling,
parallel Zeeman field, and $s$-wave superconductivity has also been applied
to electron- and hole-doped one dimensional semiconducting wires. \cite%
{Long-PRB,Roman,Oreg,Zhang-Tewari-2} The one dimensional topological systems
are particularly useful for constructing quasi-two-dimensional quantum wire
networks \cite{Alicea-Network} which can potentially be used \cite%
{Alicea-Network,Hassler-QC,Sau-Tewari-QC} as platforms for non-Abelian
statistics \cite{Nayak-Wilczek,Read-Green,Ivanov,Stern} and universal
quantum computation \cite{Kitaev-QC,Nayak-RMP} in the Bravyi-Kitaev scheme.%
\cite{Bravyi-QC} For the purpose of studying anomalous topological
transverse response functions such as anomalous Hall and Nernst effects, in
this paper we will confine ourselves to two-dimensional hole- and
electron-doped semiconductor thin films.

The substantive difference between the topological states in the  electron
and hole doped semiconductors lies in the fact that while the
superconducting order parameter in the former is similar  to the chiral $p$%
-wave ($p_x+ip_y$) type, the order parameter in the hole-doped system in a
perpendicular Zeeman field is predominantly  chiral $f$-wave ($f_x+if_y$)
type. \cite{Mao} A strong perpendicular Zeeman field, however, is difficult
to realize experimentally with a magnetic field because of the unwanted
orbital effects which are pair breaking.\cite{Alicea-Tunable} Following
Ref.~[\onlinecite{Alicea-Tunable}], in this paper we will introduce the
geometry and Hamiltonian for the hole-doped systems which can  support
topological superconductivity in the presence of an in-plane Zeeman field
making it easier to produce with a magnetic field. We will then analyze the
Berry phase mediated topological Hall and thermoelectric effects in
electron- and hole-doped systems under external  conditions necessary for
realizing topological superconductivity with broken time-reversal symmetry.
Note that, since we are considering  the spontaneous or anomalous components
of the Hall and Nernst effects, we will only treat two-dimensional systems
in the presence of an \textit{in-plane} Zeeman splitting.

In all the systems mentioned above the chemical potential regime that
corresponds to the topological state in the presence of proximity effect is
characterized by a single (or odd number of) Fermi  surface which breaks the
fermion doubling theorem.\cite{Fermion-Doubling} It has been argued that, in
the limit of vanishing superconducting pair potential $\Delta_0$,  the
single (or odd number) of Fermi surface at the chemical potential
constitutes a necessary condition for the  existence of topological
superconductivity and Majorana fermions at order parameter defects. \cite%
{Kitaev-1D} The breakdown of the fermion doubling theorem  and topological
superconductivity in all the above cases are achieved by the introduction of
a strong Zeeman splitting greater than at least  the proximity induced
superconducting pair potential $\Delta_0$. At such high values of the Zeeman
splitting the superconductivity itself survives because of the
spin-chirality induced on the semiconductor Fermi surface by a strong enough
spin-orbit coupling.\cite{Long-PRB}

The twin requirements of strong Zeeman splitting as well as spin-orbit
coupling on the semiconductor Fermi surface manifest themselves in
interesting Berry phase\cite{Berry} mediated topological effects which have
pronounced effects on the anomalous Hall and thermoelectric coefficients.
\cite{Sundaram,NiuReview,Macdonald,Xiao1,Xiao2,Zhang2,Zhang3,Niu1,Nagaosa,Fang,Lee,Niu2,Goswami,Dyrdal,Xie,QAHI}
 While these effects have been investigated previously, our focus in this
paper is near the chemical potential regime - called topological regime
hereafter - which supports topological superconductivity in the presence of $%
s$-wave proximity effect. We find that the topological regimes in both
electron- and hole-doped semiconductors are marked by well-defined plateaus
in the anomalous Hall and Nernst effects as a function of the chemical
potential. The plateaus are not quantized, however, in the sense of
quantized anomalous transport coefficients, because, even in the topological
regime, the electron or the hole-doped systems are not fully gapped. Rather,
in this regime of chemical potential they have a single (or odd number of)
Fermi surface which supports, in the presence of spin-chirality induced by
the spin-orbit coupling, the superconducting proximity effect. Even so, in
the topological regime, there is a clear separation in the momentum space
between the region where the Berry curvature is peaked (near the
`band-degeneracy' points in the absence of the Zeeman splitting) and the
region where the Fermi surface lies in the Brillouin zone. Since the Nernst
effect is strictly a Fermi surface quantity at zero temperature, the
vanishing of the overlapping region in the Brillouin zone between the Fermi
surface and the Berry curvature results in a vanishing anomalous Nernst
effect in the topological regime. The plateau in the Nernst coefficient at
vanishing magnitudes is flanked on either side of the topological regime by
well-defined peaks (of opposite signs) arising from the emergence of a
second small Fermi surface which destroys topological superconductivity but
gives rise to large peaks in the topological Nernst effect. The results
presented in this paper may be useful for experimentally deducing the
topological regimes of the chemical potential in electron and hole-doped
semiconductors supporting Majorana fermions in the presence of proximity
effect.

\section{Berry phase and topological Hall and Nernst effects}

As a particle moves adiabatically through a closed contour in its parameter
space it acquires a geometric phase known as a Berry phase. In a crystal
lattice the wavefunctions for a band are written as $\left\vert \Psi _{n}(%
\bm{k},\bm{r})\right\rangle =e^{i\bm{k}\cdot \bm{r}}\left\vert u_{n}(\bm{k},%
\bm{r})\right\rangle $ according to Bloch's theorem where $\left\vert u_{n}(%
\bm{k},\bm{r})\right\rangle $ is a Bloch function with the periodicity of
the lattice. The eigenfucntions are k dependent and the relevant paramater
space is the space defined by the crystal momentum \textbf{k}. The Berry
connection, $\bm{A}_{\bm{k}}=\left\langle u_{n}(\bm{k},\bm{r})\left\vert
i\nabla _{\bm{k}}\right\vert u_{n}(\bm{k},\bm{r})\right\rangle $ represents
the geometric phase acquired by a Bloch wave function through infinitesimal
movement in k-space and is a vector potential. In analogy to electrodynamics
the Berry curvature is defined as the curl of this potential as $\bm{\Omega}%
_{n}(\bm{k})=\nabla _{\bm{k}}\times \bm{A}_{\bm{k}}$ which is Berry phase
per unit area. The Berry curvature enters into the equations of motion of a
wavepacket and is responsible for many intrinsic transport properties. For a
system with time reversal symmetry and spatial inversion symmetry the Berry
curvature vanishes for all \textbf{k} so it is often ignored.

The charge current in the presence of an electric field \textbf{E} and a
temperature gradient $\nabla T$ can be written as $J_i=\sigma_{ij}E_j+%
\alpha_{ij}(\partial_j T)$ where $\sigma_{ij}$ and $\alpha_{ij}$ are the
electric and thermoelectric conductivity tensors. In the Hall effect a
current is applied and a magnetic field is present perpendicular to a
conducting sample. In this configuration an electric field is generated
perpendicular to the current so that off diagonal terms of $\sigma_{ij}$ are
non-zero. Similarly for the Nernst effect a current will arise normal to the
temperature gradient when a perpendicular magnetic field is present. Below
we discuss the anomalous or topological Hall and Nernst effects for a system
where there is no perpendicular magnetic field but there is still a
contribution to the Hall and Nernst effects due to the presence of a
non-trivial Berry curvature.

In the presence of an electric field \textbf{E}, the group velocity of a
Bloch electron is written as \cite{NiuReview}
\begin{equation}
\dot{\bm{r}}=\frac{1}{\hbar }\frac{\partial \epsilon _{n}(\bm{k})}{\partial %
\bm{k}}+\frac{e}{\hbar }\mathbf{{E}\times \bm{\Omega}_{n}(\bm{k})}
\end{equation}%
where the first term is the usual band dispersion and the second term is
called the anomalous velocity. This anomalous velocity is responsible for
the intrinsic contribution to the anomalous Hall and anomalous Nernst
effects, with the Berry curvature acting like a magnetic field in k-space.
With the inclusion of the anomalous velocity it immediately follows that by
summing the anomalous velocity over all occupied states the charge conductivity is
written as,\cite{Xiao2,Zhang2}

\begin{equation}
\sigma _{xy}=\frac{e^{2}}{\hbar }\sum_{n}\int \frac{dk_{x}dk_{y}}{(2\pi )^{2}%
}\Omega _{n}f(E_{n}(\bm{k}))  \label{conductivity}
\end{equation}%
where $f(E_{n})=1/(1+\text{exp}(E_{n}-\mu )/k_{B}T)$ is the Fermi
distribution function, $k_{B}$ is the Boltzmann constant and $T$ is the
temperature.

In order to write an expression for the anomalous Nernst coefficient we
first look at the coefficient $\bar{\alpha}_{xy}$ which relates the heat
current and electric field by $J_x^h=\bar{\alpha}_{xy} E_y$. $\alpha_{xy}$
can then be solved for by making use of the Onsager relation $\bar{\alpha}%
_{xy}=T \alpha_{xy}$. The transverse heat coefficient may be written as the
velocity multiplied by the entropy density,

\begin{equation}
\bar{\alpha}_{xy}=T\alpha _{xy}=\frac{e}{\beta \hbar }\sum_{n}\int \frac{%
dk_{x}dk_{y}}{(2\pi )^{2}}\Omega _{n}s_{n}(k),
\label{transverseheatcoefficient}
\end{equation}%
where the entropy density of an electron gas is given as $s(k)=-f_{k}\text{ln%
}f_{k}-(1-f_{k})\text{ln}(1-f_{k})$ with $f_{k}$ as the Fermi distribution
function. Using the above equation and this form for the entropy density,
the coefficient $\alpha _{xy}$ may be re-written as:\cite{Xiao2,Zhang2}

\begin{eqnarray}
\alpha _{xy} &=&\frac{e}{\hbar }\frac{1}{T}\sum_{n}\int \frac{dk_{x}dk_{y}}{%
(2\pi )^{2}}\Omega _{n}\times   \notag \\
&&\left\{ E_{n}(\bm{k})f(E_{n}(\bm{k}))-k_{B}T\text{log}[1-f(E_{n}(\bm{k}%
))]\right\} .  \label{alphaxy}
\end{eqnarray}

Through the use of these Berry phase mediated thermoelectric effects one can
characterize the topological regimes in chemical potentials which support
the topological superconducting state in the presence of superconducting
proximity effect. We will deliberately choose quantum well configurations
with appropriate spin-orbit couplings which will allow an in-plane Zeeman
field for the topological state so the conventional Hall and thermoelectric
effects make no contributions in experiments.

\section{Topological Hall and Nernst effects in electron doped semiconductors%
}

As a warm-up, following Alicea, \cite{Alicea-Tunable} we first consider a
zinc-blend semiconductor quantum well grown in the [110] direction with an
in-plane magnetic field applied parallel to the semiconductor film. In such
a quantum well we expect both Rashba and Dresselhaus spin orbit couplings to
be present and the Hamiltonian of this system on the $(x-y)$ plane is
written as, \cite{Alicea-Tunable}

\begin{equation}
H=\frac{\hbar^2 k^2}{2m^*}+\alpha_R(\bm{\sigma}\times \bm{k})\cdot \hat{z}%
+\alpha_D k_x \sigma_z +h_y \sigma_y  \label{2degHamiltonian}
\end{equation}

\begin{figure}[t]
\centering
\includegraphics[width=7cm]{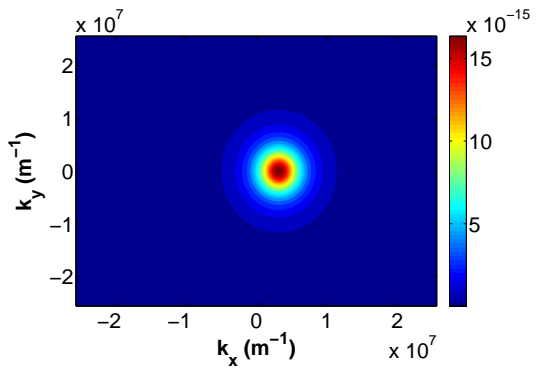}
\caption{Contour plot of the Berry curvature $\bm{\Omega}_{-}$ in the lower
band of electron-doped semiconductors. The Berry curvature is sharply peaked
at a finite value of $k_{x}$ near the band-degeneracy point in the absence
of the Dresselhaus coupling.}
\label{fig:2degBC}
\end{figure}

\begin{figure}[t]
\centering
\includegraphics[width=7cm]{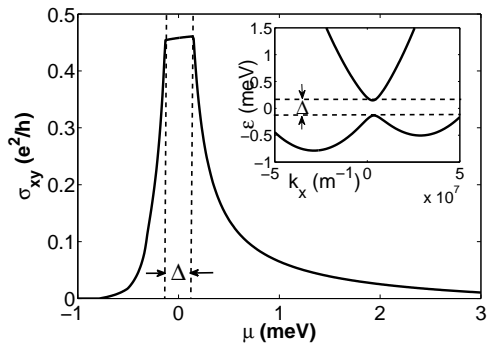}
\caption{Anomalous Hall conductivity versus chemical potential $\protect\mu$
at zero temperature for a two-dimensional electron-doped semiconductor. We
have used $\protect\alpha_R=\protect\alpha_D=4.74 \times 10 ^4$ m/s $%
m^*=.067 m_e$ and $h_y=0.2$ meV. The electronic band structure versus $k_x$
is shown in the inset along with the minimum energy gap $\Delta$ indicated
by the dashed lines}
\label{fig:2degAHE}
\end{figure}

Here, $\alpha _{R}$ and $\alpha _{D}$ are the strengths of the Rashba and
Dresselhaus couplings, $m^{\ast }$ is the effective mass of the electrons, $%
h_{y}$ the magnitude of the in-plane Zeeman field and the $\bm{\sigma}%
=(\sigma _{x},\sigma _{y},\sigma _{z})$ are the Pauli matrices.
Diagonalizing the Hamiltonian yields energy eigenvalues of $E_{\pm }=\frac{%
h^{2}k^{2}}{2m^{\ast }}+E_{0}$, where $E_{0}=\sqrt{(\alpha
_{D}k_{x})^{2}+(\alpha _{R}k_{y})^{2}+(\alpha _{R}k_{x}-h_{y})^{2}}$. The
degeneracy between the two bands is lifted by the presence of both
Dresselhaus spin orbit coupling and the in-plane magnetic field. It is easy
to see that, in the absence of the Dresselhaus coupling, the in-plane Zeeman
splitting can be reabsorbed in the Hamiltonian by a re-definition of the
momentum $\alpha _{R}k_{x}\rightarrow \alpha _{R}k_{x}+h_{y}$ which leaves
the system gapless even with the Zeeman splitting. The existence of a
non-zero Dresselhaus term $\alpha _{D}$ ensures that a finite gap is created
near the band-degeneracy points at a non-zero value of $k_{x}$ even after
this re-definition. The minimum gap between the bands, $\Delta =2\alpha
_{D}h_{y}/\sqrt{\alpha _{R}^{2}+\alpha _{D}^{2}}$, is located at $%
k_{x}=\alpha _{R}h_{y}/(\alpha _{R}^{2}+\alpha _{D}^{2})$ and $k_{y}=0$,
\cite{QAHI} which is shown in the inset of figure \ref{fig:2degAHE}.

We calculate the Berry curvatures for this system through $\bm{\Omega}_{\pm
}=2\text{Im}\left\langle \frac{\partial \Phi _{\pm }}{\partial k_{x}}%
\left\vert \frac{\partial \Phi _{\pm }}{\partial k_{y}}\right. \right\rangle
\hat{z}$, where $\Phi \pm $ are the eigenstates of the Hamiltonian in Eq.~(%
\ref{2degHamiltonian}). Evaluating this expression analytically we find that%
\cite{QAHI}
\begin{equation}
\bm{\Omega}_{\pm }=\mp \frac{\alpha _{R}\alpha _{D}h_{y}}{2E_{0}^{3}}\hat{z},
\label{2degBC}
\end{equation}%
which are of equal magnitude and opposite signs in the two bands. In Fig.~%
\ref{fig:2degBC} we plot $\bm{\Omega_-}$ where it can be seen that the Berry
curvature is sharply peaked at the gap minimum between the two bands, i.e.,
at the band degeneracy point in k-space in the absence of the Dresselhaus
coupling. We note here that the Berry curvatures are only non-zero for
non-zero values of $\alpha _{D},\alpha _{R}$ and $h_{y}$.

Figure \ref{fig:2degAHE} shows the dependence of the anomalous Hall
conductivity on the chemical potential $\mu$ at zero temperature. The band
structure near the origin is shown in the inset. Below $\mu\ \cong -0.5meV$
there is no contribution to the integral for the anomalous Hall coefficient
(see Eq.~(\ref{conductivity})) as the states near the band gap minimum which
have large Berry curvatures (Fig \ref{fig:2degBC}) are unoccupied. As $\mu$
increases these states are filled and there is a positive contribution to
the integral from $\Omega_-$. The quasi-plateau in the hall conductivity for
$\mu$ corresponds to the energy gap between the two bands where there is a
single large Fermi surface away from the origin. It is in this regime of
chemical potential that the system supports the topological superconducting
state because of the breakdown of the fermion doubling theorem. \cite%
{Sau1,Index-Theorem,Long-PRB,Alicea-Tunable} At higher $\mu$ the upper band
becomes occupied, leading to a cancelation of the anomalous Hall
conductivity due to the equal magnitude but opposite sign of the Berry
curvatures of the two bands.

\begin{figure}[t]
\centering
\includegraphics[width=7cm]{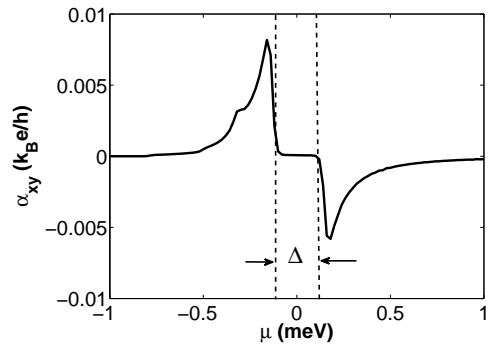}
\caption{Anomalous Nernst coefficient versus chemical potential for 2DEG.
The dashed lines show the regime of the energy gap $\Delta$ between the
bands. The same parameters have been used as in Fig \protect\ref{fig:2degAHE}
with the exception of $T=0.1K$}
\label{fig:2degANE}
\end{figure}

The anomalous Nernst coefficient for an electron-doped semiconductor near
the topological regime is plotted against $\mu$ in Fig.~\ref{fig:2degANE}.
At low temperatures the entropy density $s_n(k)$ is sharply peaked at the
Fermi surface, such that the integrand in Eq.~(\ref{alphaxy}) is non-zero
only for values of $\mu$ corresponding to the intersection of the Fermi
surface(s) and the states near the minimum band gap (close to the origin)
where the Berry curvatures are non-zero. Figure \ref{fig:2degANE}
illustrates this behavior with a positive peak from the lower band and a
negative contribution from the upper band. For the chemical potential
slightly below or above the topological regime the system has two (or an
even number of) Fermi surfaces one of which lies close to the band
degeneracy point. This small Fermi surface, because of a non-zero Berry
curvature, gives a finite contribution to the anomalous Nernst effect which
differs in sign between the regimes below and above the topological regime.
As the topological regime is approached from either side, the Berry
curvatures become increasingly sharply peaked (because of a decreasing $E_0$%
, see Eq.~(\ref{2degBC})) while the Fermi surface areas themselves go down
resulting in a pair of peaks of opposite signs surrounding the topological
regime. The topological regime itself is characterized by a single (or odd
number of) Fermi surface with no weight near the band degeneracy points with
significant Berry curvatures. Thus, the plateau between the two peaks in
Fig.~\ref{fig:2degANE} corresponds to the minimum gap separating the energy
bands and indicates the regime in which the topological superconducting
state is possible.

\section{Topological Hall and Nernst effects in hole doped semiconductors}

Next we consider thin film hole doped semiconductor quantum wells grown in
the [110] direction. In zinc-blende semiconductors the band structure is
described by $\bm{k}\cdot \bm{p}$ perturbation theory.\cite{Winkler} Here
the top valence bands consist of three $p$-orbitals which have an angular
momentum $L=1$ leading to a six fold degeneracy at the origin. Including
spin, the total angular momentum operator becomes $J=L+S$ so that there are
four $J=3/2$ (heavy hole and light hole) and two (split-off) $J=1/2$ bands
which are separated by a large energy gap through atomic spin orbit
coupling. We focus on the $J=3/2$ bands described by the Luttinger
Hamiltonian near $k=0$ which is written as,
\begin{equation}
H_{L}=\frac{1}{m}%
\begin{bmatrix}
P+Q & -S & R & 0 \\
-S^{\ast } & P-Q & 0 & R \\
R^{\ast } & 0 & P-Q & S \\
0 & R^{\ast } & S^{\ast } & P+Q%
\end{bmatrix}
\label{HL}
\end{equation}%
where the quantities $P,Q,R$ are functions of the Luttinger parameters $%
\gamma _{1},\gamma _{2},\gamma _{3}$ and the momentum components $%
k_{x},k_{y},k_{z}$ and act on the basis $\left\vert j,m\right\rangle $ with $%
j=3/2$ and $m=3/2,1/2,-1/2,-3/2$ with spin quantization in the growth
direction. In Ref.~[\onlinecite{Mao}] topological superconducting states
with a chiral-$f$-wave symmetry have been shown to exist in hole-doped
quantum wells grown in the [001] direction in the presence of a
perpendicular Zeeman field and Rashba spin-orbit coupling. A perpendicular
Zeeman field, however, is unsuitable for investigations of the anomalous
Hall and Nernst effects because such a magnetic field itself gives rise to
conventional Hall and Nernst effects which are expected to dominate over the
Berry-phase mediated anomalous response. Therefore, in order to uncover the
anomalous Hall and Nernst effects in the topological regime of the
hole-doped systems, we consider the semiconductor quantum well in the [110]
direction with a Dresselhaus spin-orbit coupling and a parallel Zeeman field.

For the [110] growth direction, we have for the functions $P,Q,R$, \cite%
{PhysRevB.52.11132}
\begin{eqnarray}
P & = & \frac{1}{2} \gamma_1 (k_x^2+k_y^2+k_z^2) \\
Q & = & \frac{1}{2} \gamma_2 (k_x^2-\frac{1}{2}k_y^2-\frac{1}{2}k_z^2)+\frac{%
3}{4}\gamma_3(k_y^2-k_z^2)  \notag \\
S & = & \sqrt{3} (\gamma_3 k_x - i \gamma_2 k_y) k_z  \notag \\
R & = & \frac{\sqrt{3}}{4}((\gamma_2+\gamma_3)
k_y^2+(\gamma_2-\gamma_3)k_z^3)  \notag \\
&& -\frac{\sqrt{3}}{2}(\gamma_2 k_x^2-2 i \gamma_3 k_x k_y )  \notag
\label{coefficients}
\end{eqnarray}
\begin{figure}[t]
\centering
\includegraphics[width=8cm]{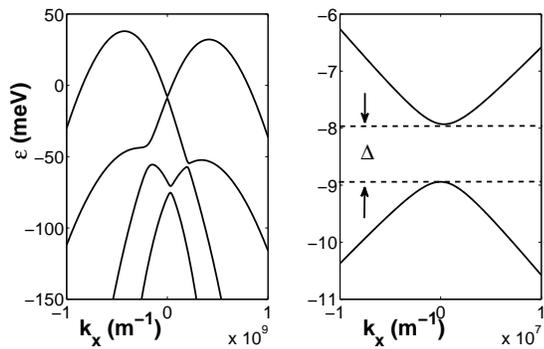}
\caption{(a) Band structure of the 2DHG in the $k_x$ direction for $k_y=0$.
(b) Energy gap at origin between lower bands. Here we have used $\protect%
\alpha_R=\protect\alpha_D=2 \times 10 ^5$ m/s, $h_y=5$ meV, $\protect\gamma%
_1=6.92$,$\protect\gamma_2=2.1$, $\protect\gamma_3=2.9$ and $a=8nm$ which
are chosen for GaAs.}
\label{fig:2DHGbands}
\end{figure}
where we now have $k_x$, $k_y$, $k_z$ in the $[0,0,-1]$, $[-1,1,0]$, and $%
[1,1,0]$ directions, respectively. Due to the confinement of the quantum
well, the momentum is quantized in the growth direction and is approximated
by $\left\langle k_z\right\rangle=0$ and $\left\langle
k_z^2\right\rangle\approx (\pi /a)^2$ where $a$ is the width of the well.
This confinement projects the above Hamiltonian from three dimensions into
two and also serves to lift the degeneracy between the heavy and light hole
bands.

The single particle Hamiltonian for a two dimensional hole gas in a $[110]$
quantum well which is expected to support topological superconductivity with
$s$-wave proximity effect is the sum of the Luttinger, spin-3/2 Rashba, and
Dresselhaus terms,
\begin{equation}
H=H_{L}+\alpha _{R}(\mathbf{J}\times \bm{k})\cdot \hat{z}+\alpha
_{D}k_{x}J_{z}+h_{y}J_{y}  \label{2dhg hamiltonian}
\end{equation}%
where \textbf{J} is the total angular momentum operator given by the spin
3/2 matrices, $\gamma _{1}$, $\gamma _{2}$ and $\gamma _{3}$ are the
Luttinger parameters, $\alpha _{R}$ and $\alpha _{D}$ are the strengths of
the Rashba and Dresselhaus couplings and $h_{y}$ is an in-plane Zeeman
splitting as before. The form of the Rashba and Dresselhaus couplings in Eq.
(\ref{2dhg hamiltonian}) for holes can be derived using nearly-degenerate
pertubation theory.\cite{Winkler,Mao3} The band structure for this
Hamiltonian is illustrated in Fig.~\ref{fig:2DHGbands} with parameters
chosen for GaAs. It is clear from Fig.~\ref{fig:2DHGbands} that the combined
effects of the Rashba, Dresselhaus, and the in-plane Zeeman splitting give
rise to several regimes of the chemical potential where a spectral gap opens
up near the band-degeneracy points. In analogy with the electron-doped
semiconductors, when the chemical potential falls within the spectral gaps
(topological regime) the system has an odd number of Fermi surfaces leading
to topological superconductivity in the [110] grown hole-doped well in the
presence of superconducting proximity effect.

\begin{figure}[t]
\centering
\includegraphics[width=7cm]{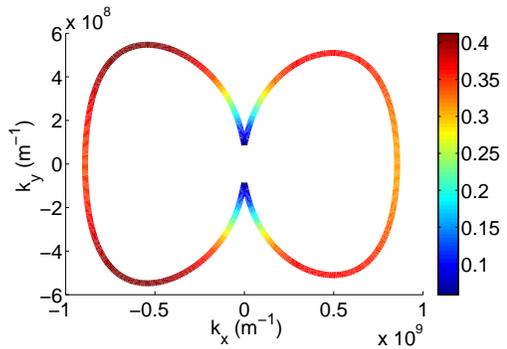}
\caption{(Color online) Contour plot of the Fermi surface at $\protect\mu$
corresponding to the topological regime ($\protect\mu=-8.5$ meV). Color plot
indicates the proportion of light-hole character ($m_j=\pm1/2$) in the top
valence band.}
\label{fig:probdensity}
\end{figure}

Next, for a robust $s$-wave proximity effect on a hole-doped quantum
well we need to ensure that the top valence band orbital wave functions
couple with the orbitals of the adjacent $s$-wave superconductor. That this
coupling is not automatically assured can be seen from the fact that the
valence band holes are generically $p$-wave (in contrast to the conduction
band electrons which are typically $s$-wave), and therefore coupling with
the $s$-orbitals of the superconductor puts certain constraints on the value
of $m$ of the top valence band wavefunctions. To illustrate this, suppose
the top valence band quantum state contains contributions only from pure
eigenstates of $J_z$ as $\left|j=\frac{3}{2},m=\pm\frac{3}{2}\right\rangle$.
Since $m=m_l+m_s$ where $m_l$ and $m_s$ are orbital and spin angular
momentum quantum numbers, respectively, $m=\pm \frac{3}{2}$ implies that
this state must consist of only $m_l=\pm 1$, $m_s=\pm \frac{1}{2}$ states.
However in this case there will be no overlap between the superconductor
orbitals if they are $s$-wave ($m_l=0$) and the $m_l=\pm1$ orbitals of the
valence band, and no proximity induced superconductivity can be induced in
the valence band. Therefore unlike in the electron conduction band, which
consists of $s$ orbitals, we must investigate the orbital angular momentum
character of the valence bands for holes. 

The eigenstates of the Hamiltonian
in Eq.~(\ref{2dhg hamiltonian}) are pure eigenstates of the operator $J_z$
only at $k=0$, but as $k$ increases the bands become a mixture $J_z$
eigenstates and can be written as a linear combination in the form $%
\left|\Phi_n\right\rangle=c_1(\bm{k}) \left|\frac{3}{2},\frac{3}{2}%
\right\rangle+c_2(\bm{k})\left|\frac{3}{2},\frac{1}{2}\right\rangle+ c_3(%
\bm{k})\left|\frac{3}{2},\frac{-1}{2}\right\rangle+c_4(\bm{k})\left|\frac{3}{%
2},\frac{-3}{2}\right\rangle$. Figure \ref{fig:probdensity} shows the Fermi
surface for a value of $\mu$ ($=-8.5$ meV) corresponding to the gap between
the HH bands (topological regime, see Fig.~\ref{fig:2DHGbands}) and the
mixing of $m_j=\pm 1/2$ states given by $\left|c_2\right|^2+\left|c_3%
\right|^2$ in the top band at the Fermi surface. Note that $m_j=\pm 1/2$
eigenstates can be re-written in the basis $\left|m_l,m_s\right\rangle$ as a
linear combination of $\left|m_l=0,1,m_s=\frac{1}{2},-\frac{1}{2}%
\right\rangle$ which guarantees the presence of $m_l=0$ states at the Fermi
surface allowing for robust proximity induced superconductivity.

Next we wish to calculate the Berry curvatures associated with the hole band
structure shown in Fig.~\ref{fig:2DHGbands} numerically. In order to
facilitate the calculations of the Berry curvatures, we use the following
expression for the Berry curvature \cite{NiuReview} which is equivalent to
the form discussed earlier before Eq.~(\ref{2degBC}),
\begin{equation}
\Omega^n_{xy}=i \sum_{n^{\prime }\neq n}\frac{\left\langle \Phi_n\left|\frac{%
\partial H}{\partial k_x}\right|\Phi_n^{\prime }\right\rangle \left\langle
\Phi_n^{\prime }\left|\frac{\partial H}{\partial k_y}\right|\Phi_n\right%
\rangle-(k_x\leftrightarrow k_y)}{(E_{n}-E_{n^{\prime }})^2}  \label{BCsum}
\end{equation}
\begin{figure}[t]
\centering
\includegraphics[width=9cm]{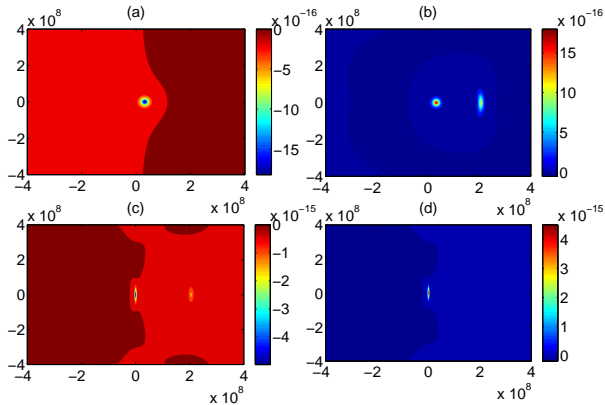}
\caption{(Color online) Contour plot of Berry curvature of each band (units
of $m^2$) where (a) corresponds to the lowest band and (d) to the highest.
The y and x axis are $k_y$ and $k_x$ respectively with units of $m^{-1}$. As
seen from the denominator of Eq.~(\protect\ref{BCsum}) these functions are
sharply peaked at values of \textbf{k} near the near-band-degeneracy points.}
\label{fig:2dhgBC}
\end{figure}
That this form of the Berry curvatures is equivalent to the earlier
expression discussed before Eq.~(\ref{2degBC}) can be understood by noting
that $\left\langle \Phi_n \left| \nabla_k \right| \Phi_{n^{\prime }}
\right\rangle(E_{n^{\prime }}-E_n)=\left\langle \Phi_n\left|\nabla_k H(k)
\right| \Phi_{n^{\prime }}\right\rangle$. \cite{NiuReview} Eq.~(\ref{BCsum})
has the additional benefit that the arbitrary phase factors of the
eigenstates from numerical diagonalization are ignored as there is no
differentiation of the eigenstates involved. A contour plot of the Berry
curvatures corresponding to each of the four bands is given in Fig. \ref%
{fig:2dhgBC}. As can be seen from this figure, the Berry curvatures are
sharply peaked at points in k-space corresponding to the minimum energy gaps
in the band structure of the holes, that is, near the near-band-degeneracy
points.

We now calculate the anomalous Hall conductivity through Eq.(\ref%
{conductivity}) where we use Eq.~(\ref{BCsum}) for the Berry curvatures,
which is equivalent to the Kubo formula in linear response theory. Fig. \ref%
{fig:2dhgAHE} shows the dependence of $\sigma_{xy}$ on $\mu$ at zero
temperature. The physics of this effect is the same as that of the electron
doped case but with Berry curvatures and a band structure that are more
complicated. Decreasing $\mu$ excites more holes, filling each band, such
that there are contributions to $\sigma_{xy}$ corresponding to the overlap
of the Fermi distribution function and Berry curvature for each band(Fig \ref%
{fig:2dhgBC}). As $\mu$ is made increasingly large and all bands are filled
the sum of the contributions approaches zero. There is again, like in the
case of the electron doped semiconductors, a small quasi-plateau
corresponding to the topological regime of the chemical potential separating
the top two bands.

\begin{figure}[t]
\centering
\includegraphics[width=7cm]{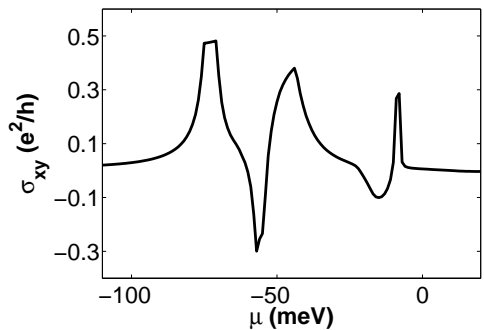}
\caption{Plot of the anomalous Hall conductivity for the hole doped
semiconductor at T=0 with the same paramaters as those used in Fig. \protect
\ref{fig:2DHGbands}.}
\label{fig:2dhgAHE}
\end{figure}

\begin{figure}[t]
\centering
\includegraphics[width=7cm]{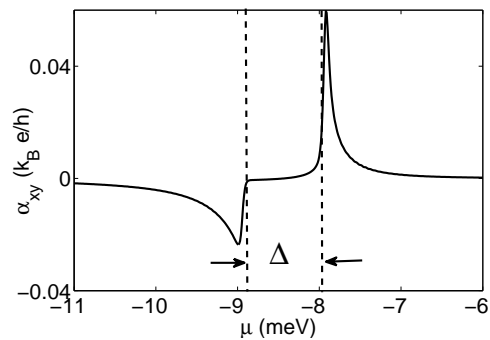}
\caption{Plot of the anomalous Nernst coefficient for the $[110]$ grown
hole-doped quantum well near the topological regime of the chemical
potential. The topological regime of $\protect\mu$ is characterized by a
plateau of the anomalous Nernst coefficient (at vanishing values) flanked by
two peaks of opposite signs as in the case of the electron-doped
semiconductors. The parameters used are the same as in Fig. \protect\ref%
{fig:2DHGbands} except that $T=0.1K$.}
\label{fig:2dhgANE}
\end{figure}

The regime of chemical potential suitable for topological superconductivity
in the presence of $s$-wave proximity effect can be more clearly seen in the
anomalous Nernst coefficient which has a well-defined plateau at vanishing
values in the topological regime. We calculate the anomalous Nernst
coefficient through Eq.~(\ref{alphaxy}) with the Berry curvatures found by
using Eq.~(\ref{BCsum}). The $\mu$ dependence of the Nernst coefficient is
shown in Fig.~\ref{fig:2dhgANE} near the regime of $\mu$ near the energy gap
$\Delta$ between the top bands as illustrated in Fig.~\ref{fig:2DHGbands}.
Similar to the electron doped systems, the integrand of Eq.~(\ref{alphaxy})
is a product of the entropy density, sharply peaked at the Fermi surface,
and the Berry curvatures shown in Fig \ref{fig:2dhgBC} which are peaked near
the near-band-degeneracy points. There are two contributions to Fig. \ref%
{fig:2dhgANE} as the Fermi surface corresponding to each band sweeps through
the area of k-space near the origin where the Berry curvature is sharply
peaked. The plateau for which the coefficient $\alpha_{xy}=0$ corresponds to
the energy gap between the bands which supports the topological
superconducting state and is surrounded by well-defined peaks of opposite
sign on either side. The vanishing of the Nernst effect, as before,
originates from the clear momentum space separation between the single (or
odd number of) Fermi surface, which is a requirement for the topological
superconductivity with Majorana fermions, and the regions in the momentum
space (the near-band-degeneracy points) where the topological Berry phase is
sharply peaked.

\section{\textbf{Summary and conclusion}}

In this paper we study the intrinsic contributions to the anomalous Hall and
thermoelectric coefficients for thin film electron- and hole-doped
semiconductors with Rashba and Dresselhaus spin orbit couplings and a
suitably directed Zeeman field. Due to the presence of the spin orbit
interactions and Zeeman field, a gap is induced in both the conduction and
valence bands. When the chemical potential is inside the gap, the so called
topological regime, it has been proposed that a topological superconducting
state with Majorana fermions may be supported in the presence of $s$-wave
superconducting proximity effect. For the study of anomalous Hall and Nernst
effects, we require the applied Zeeman field to be parallel to the planes of
the semiconductor. To achieve this, we first introduce the Hamiltonian of a $%
[110]$ grown hole-doped quantum well and show that in the presence of a
parallel Zeeman field several topological regimes of chemical potential open
up which can potentially support topological superconductivity in the
presence of proximity effect. We then discuss the wave function of the top
hole-doped valence band and show that there is a considerable mixing of $%
m_{j}=\pm 1/2$ states which is necessary for proximity induced $s$-wave
superconductivity. With the Hamiltonians for the electron- as well as
hole-doped systems capable of supporting topological regimes of the chemical
potential with only in-plane magnetic fields, we discuss the associated
Berry curvatures. Time reversal and spatial inversion symmetry breaking give
rise to non-trivial Berry curvatures at the points in k-space corresponding
to local minimum gaps between energy bands, the so-called
near-band-degeneracy points. We make use of this fact to show that the
topological regimes of the chemical potential generically have well-defined
plateaus in both anomalous Hall and Nernst effects. While the plateau in the
anomalous Hall coefficient is at a non-zero value, the Nernst coefficient
saturates in the topological regime at $\alpha _{xy}=0$. The plateau at $%
\alpha _{xy}=0$ is surrounded by well-defined peaks of the anomalous Nernst
effect of opposite signs indicating the emergence of the topological regime.
The vanishing of the Nernst effect in the topological regime originates from
the clear momentum space separation between the single (or odd number of)
Fermi surface, a requirement for the topological superconductivity with
Majorana fermions, and the regions in the momentum space (the
near-band-degeneracy points) where the topological Berry phase is sharply
peaked.

\section{Acknowledgement}
This work is supported by DARPA-MTO (FA9550-10-1-0497), NSF (PHY-1104527), DARPA-YFA (N66001-10-1-4025), and NSF (PHY-1104546).


\begin{thebibliography}{99}


\bibitem{Schnyder} A. P. Schnyder, S. Ryu, A. Furusaki, and A. W. W. Lud-
wig, Phys. Rev. B \textbf{78} 195125 (2008); A. P. Schnyder, S. Ryu, A.
Furusaki, and A. W. W. Ludwig, AIP Conf. Proc. \textbf{1134} 10 (2009).

\bibitem{Kitaev-AIP} A. Yu Kitaev AIP Conf. Proc. \textbf{1134} 22 (2009).

\bibitem{Ruy} S. Ruy, A. Schnyder, A. Furusaki, A. W. W. Ludwig, New J.
Phys. \textbf{12}, 065010 (2010).

\bibitem{Majorana} E. Majorana, Nuovo Cimento \textbf{5}, 171 (1937).

\bibitem{Wilczek} F. Wilczek, Nature Physics \textbf{5}, 614 (2009).

\bibitem{Nayak-Wilczek} C. Nayak, and F. Wilczek, Nucl. Phys. B \textbf{479}%
, 529 (1996).

\bibitem{Read-Green} N. Read and D. Green, Phys. Rev. B \textbf{61}, 10267
(2000).

\bibitem{Ivanov} D. A. Ivanov, Phys. Rev. Lett. \textbf{86}, 268 (2001).

\bibitem{Stern} A. Stern, F. von Oppen, E. Mariani, Phys. Rev. B \textbf{70}%
, 205338 (2004).

\bibitem{Kitaev-QC} A. Kitaev, Ann. Phys. \textbf{303}, 2 (2003).

\bibitem{Nayak-RMP} C. Nayak, S. H. Simon, A. Stern, M. Freedman, S. Das
Sarma, Rev. Mod. Phys. \textbf{80}, 1083 (2008).

\bibitem{Sau1} J. D. Sau, R. M. Lutchyn, S. Tewari, S. Das Sarma, Phys. Rev.
Lett. \textbf{104}, 040502 (2010).

\bibitem{Index-Theorem} S. Tewari, J. D. Sau, S. Das Sarma, Annals of
Physics \textbf{325}, 219, (2010).

\bibitem{Long-PRB} J. D. Sau, S. Tewari, R. Lutchyn, T. Stanescu and S. Das
Sarma, Phys. Rev. B \textbf{82}, 214509 (2010).

\bibitem{Zhang-Tewari} C. Zhang, S. Tewari, R. M. Lutchyn, S. Das Sarma,
Phys. Rev. Lett. \textbf{101}, 160401 (2008).

\bibitem{Sato-Fujimoto} M. Sato, Y. Takahashi, S. Fujimoto, Phys. Rev. Lett.
\textbf{103}, 020401 (2009).

\bibitem{Alicea-Tunable} J. Alicea, Phys. Rev. B \textbf{81}, 125318 (2010).

\bibitem{Mao} L. Mao, J. Shi, Q. Niu, C. Zhang, Phys. Rev. Lett. \textbf{106}%
, 157003 (2011).

\bibitem{Roman} R. M. Lutchyn, J. D. Sau, S. Das Sarma, Phys. Rev. Lett.
\textbf{105}, 077001 (2010) .

\bibitem{Oreg} Y. Oreg, G. Refael, F. V. Oppen, Phys. Rev. Lett. \textbf{105}%
, 177002 (2010).

\bibitem{Zhang-Tewari-2} L. Mao, M. Gong, E. Dumitrescu, S. Tewari, C.
Zhang, Phys. Rev. Lett. (in press); arXiv:1105.3483.

\bibitem{Alicea-Network} J. Alicea, Y. Oreg, G. Refael, F. von Oppen, M. P.
A. Fisher, Nature Physics \textbf{7}, 412-417 (2011).

\bibitem{Hassler-QC} F. Hassler, A. R. Akhmerov, C.-Y Hou, C. W. J.
Beenakker, New J. Phys. \textbf{12}, 125002 (2010).

\bibitem{Sau-Tewari-QC} J. D. Sau, S. Tewari, S. Das Sarma, Phys. Rev. A
\textbf{82}, 052322 (2010).

\bibitem{Bravyi-QC} S. Bravyi and A. Kitaev, Phys. Rev. A \textbf{71},
022316 (2005).

\bibitem{Fermion-Doubling} H. Nielssen and N. Ninomiya, Phys. Lett. \textbf{%
130B}, 389 (1983).

\bibitem{Kitaev-1D} A. Y. Kitaev, Physics-Uspekhi \textbf{44}, 131 (2001).

\bibitem{Berry} M. V. Berry, Proc. R. Soc. London, Ser. A \textbf{392}, 45
(1984).

\bibitem{Sundaram} G. Sundaram and Q. Niu, Phys. Rev. B \textbf{59}, 14915
(1999).

\bibitem{NiuReview} D. Xiao, M.-C. Chang, Q. Niu, Rev. Mod. Phys. \textbf{82}%
, 1959 (2010).

\bibitem{Macdonald} N. Nagaosa, J. Sinova, S. Onoda, A. H. MacDonald, and N.
P. Ong, Rev. Mod. Phys. \textbf{82}, 1539 (2010).

\bibitem{Xiao1} D. Xiao, W. Yao, and Q. Niu, Phys. Rev. Lett. \textbf{99},
236809 (2007).

\bibitem{Xiao2} D. Xiao, Y. Yao, Z. Fang, and Q. Niu, Phys. Rev. Lett.
\textbf{97}, 026603 (2006).

\bibitem{Zhang2} C. Zhang, S. Tewari, V. M. Yakovenko, and S. Das Sarma,
Phys. Rev. B \textbf{78}, 174508 (2008).

\bibitem{Zhang3} C. Zhang, S. Tewari, S. Das Sarma, Phys. Rev. B \textbf{79}%
, 245424 (2009).

\bibitem{Niu1} T. Jungwirth, Q. Niu, and A. H. MacDonald, Phys. Rev. Lett.
88, 207208 (2002).

\bibitem{Nagaosa} S. Murakami, N. Nagaosa, and S.-C. Zhang, Science \textbf{%
301},1348 (2003).

\bibitem{Fang} Z. Fang, N. Nagaosa, K. S. Takahashi, A. Asamitsu, R.
Mathieu, T. Ogasawara, H. Yamada, M. Kawasaki, Y. Tokura, and K. Terakura,
Science 302, 92 (2003).

\bibitem{Lee} W.-L. Lee, S. Watauchi, V. L. Miller, R. J. Cava, and N. P.
Ong, Phys. Rev. Lett. \textbf{93}, 226601 (2004).

\bibitem{Niu2} J. Shi, G. Vignale, D. Xiao, and Q. Niu, Phys. Rev. Lett.
\textbf{99}, 197202 (2007).

\bibitem{Goswami} S. Goswami, C. Siegert, M. Pepper, I. Farrer, D. A.
Ritchie, and A. Ghosh, Phys. Rev. B 83, 073302 (2011).

\bibitem{Dyrdal} A. Dyrdal, and J. Barnas, arXiv:1104.3036.

\bibitem{Xie} S.-G. Cheng, Y. Xing, Q.-F. Sun, and X. C. Xie, Phys. Rev. B
\textbf{78}, 045302 (2008).

\bibitem{QAHI} Y. Zhang, C. Zhang, Phys. Rev. B \textbf{84}, 085123 (2011).



\bibitem{Winkler} R. Winkler, Spin-Orbit Coupling Effects in Two-Dimensional
Electron and Hole Systems (Springer, New York, 2003).



\bibitem{PhysRevB.52.11132} G. Fishman, Phys. Rev. B \textbf{52}, 11132
(1995).

\bibitem{Mao3} L. Mao, M. Gong, S. Tewari, and C. Zhang, to be published.
\end{thebibliography}
\end{document}